\documentclass[aps,prl,unsortedaddress,superscriptaddress,showpacs,nofootinbib,10pt]{revtex4-2} 
\usepackage{graphicx} 
\usepackage{verbatim} 
\usepackage{amsmath}
\usepackage{array}
\usepackage{booktabs}
\usepackage{epstopdf}
\usepackage{epsfig}

\begin{document}

\title{Status and prospects of the $\chi_{c1}(3872)$ study at BESIII}

\author{Hongjian Zhou}
\affiliation{Department of Physics, Nanjing Normal University, Nanjing 210023, China\\}

\author{Xin Liu}
\affiliation{Department of physics and electronic technology,
Liaoning Normal University, Dalian 116029, China\\}

\author{Yueming Zhang}
\affiliation{Department of physics and electronic technology,
Liaoning Normal University, Dalian 116029, China\\}

\author{Chunhua Li}
\email{chunhua1319@126.com}
\affiliation{Department of Physics, Nanjing Normal University, Nanjing 210023, China\\}
\affiliation{Department of physics and electronic technology, Liaoning Normal University, Dalian 116029, China\\}


\begin{abstract}
The $\chi_{c1}(3872)$ serves as a pivotal role for understanding hadronic structures, remaining one of the most extensively studied exotic particles despite the experimental discovery of numerous unconventional hadronic states. 
Sustained experimental and theoretical investigations into the 
particle over the past two decades have propelled its study into a high-precision regime, marked by refined measurements of its decay dynamics and line shape, thereby offering critical insights to resolve longstanding debates between molecular, tetraquark, hybrid, and charmonium interpretations of this particle.
The BESIII experiment has made seminal contributions to the study of the 
$\chi_{c1}(3872)$, leveraging its unique capabilities in high-statistics data acquisition and low-background condition. This article gives a concise review and prospects of the study of the $\chi_{c1}(3872)$ from the BESIII experiment.
\end{abstract}

\maketitle

\section{Introduction}

The $\chi_{c1}(3872)$, first observed in 2003 by the Belle collaboration [1], remains the most enigmatic state among the charmonium-like XYZ particles two decades after its discovery. 
The precise determination of the $\chi_{c1}(3872)$ mass and width is critical for understanding its nature, given its striking proximity to the \(D^{*0}\bar{D}^0\) mass threshold at \(3871.69 \pm 0.07\ \text{MeV}\). Multiple experiments have employed Breit-Wigner parametrizations to analyze the resonant parameters in the decay channels containing \(J/\psi\). The Particle Data Group (PDG) reports the global averages of the mass and width to be \(M = 3871.64 \pm 0.06\ \text{MeV}\) and \(\Gamma = 1.19 \pm 0.21\ \text{MeV}\)~\cite{pdg}. However, the measurements through the open-charm channel \(B \to D^{*0}\bar{D}^0K\) decays by Belle and BaBar yield systematically higher values (\(M > 3873\ \text{MeV}\), \(\Gamma > 3\ \text{MeV}\))~\cite{Belle_dm,BaBar_dm}, significantly exceeding those from the hidden-charm final states. This discrepancy in lineshapes between the hidden- and open-charm decay modes of 
the $\chi_{c1}(3872)$ likely arises from coupled-channel effects inducing threshold distortions near \(D^{*0}\bar{D}^0\), rendering symmetric Breit-Wigner descriptions inadequate in the \(D^{*0}\bar{D}^0\) spectrum.
To address the coupled-channel distortions, Flatté-model parametrization is applied to 
describe the line shape.  
Refs.~\cite{flat1,flat2} implemented Flatté-model analyses of Belle and BaBar data for \(X(3872) \to \pi^+\pi^- J/\psi\), \(\pi^+\pi^-\pi^0 J/\psi\), and \(D^{*0}\bar{D}^0\), revealing significant \(\chi_{c1}(2P)\) admixture in Belle datasets. 
Subsequent Flatté analyses by LHCb, Belle, and BESIII~\cite{flat_lhcb,Belle_dm,flat_bes} remain inconclusive due to limited statistics, particularly in the \(\chi_{c1}(3872) \to D^{*0}\bar{D}^0\) channel where signal yields are severely constrained. Substantial improvements in \(D^{*0}\bar{D}^0\) event statistics are therefore imperative for precision line shape studies to disentangle the \(\chi_{c1}(3872)\) internal structure. 

Over the past two decades, extensive measurements have revealed multiple decay channels, including \(\chi_{c1}(3872) \rightarrow \pi^+\pi^- J/\psi\), \(D^{*0}\bar{D}^0\), \(\omega J/\psi\), \(\gamma J/\psi\), \(\gamma \psi(2S)\), and \(\pi^0\chi_{c1}\)~\cite{Be2011a, Ba2008a, Be2010c, Ba2008c, Ba2010d, Bs2, Bs3, Be2011b, Ba2009b}. 
Among its decay modes, the radiative transitions \(\chi_{c1}(3872) \rightarrow \gamma J/\psi\) and \(\gamma \psi(2S)\) are particularly interesting. These channels exhibit stark contrasts in branching fractions between conventional charmonium and exotic configurations. For example, the decay width for \(\gamma \psi(2S)\) is significantly enhanced in conventional charmonium compared to the molecular scenarios. This difference provides critical leverage for elucidating the \(\chi_{c1}(3872)\) composition.
Nevertheless, consensus regarding its nature remains elusive. The theoretical interpretations proposals include
a \(D^{*0}\bar{D}^0\) molecular configuration~\cite{th1}, 
a state mixing molecular \(D^{*0}\bar{D}^0\) components with conventional \(\chi_{c1}(2P)\) charmonium~\cite{th2}, a compact tetraquark state~\cite{th3}, and a threshold cusp at the \(D^{*0}\bar{D}^0\) mass limit~\cite{th4}.

Since the hadronic transition between $\chi_{c1}(3872)$
and $J/\psi$ is via $\chi_{c1}(3872)\to\rho^0 J/\psi$,
then there must be the charged 
$\chi_{c1}(3872)$ which can decay to $J/\psi$ via
$\chi_{c1}(3872)^\pm\to\rho^\pm J/\psi$
if the $X(3872)$ and its decay obey the isospin symmetry.
In addition, some tetraquark models
propose $\chi_{c1}(3872)$ is a four-quark bound state, and 
further predict the existence of the charged $\chi_{c1}(3872)$~\cite{fqm1,fqm2,fqm3}.
BaBar and Belle both
reported the search of the charged $\chi_{c1}(3872)$
via the $B$ meson decay $B\to X^\pm K$~\cite{cxb1,cxb2}.  
BaBar provided the upper limits 
$\mathcal{B}(B^-\to X^- \bar{K}^0, X^-\to J/\psi\pi^-\pi^0) < 22\times10^{-6}$
and 
$\mathcal{B}(B^0\to X^- K^+, X^-\to J/\psi\pi^-\pi^0) < 5.4\times10^{-6}$.
Belle provided more strict upper limits
$\mathcal{B}(B^+\to X^+ K^0)\times\mathcal{B}(X^+\to J/\psi\rho^+) < 6.1\times10^{-6}$
and 
$\mathcal{B}(\bar{B}^0\to X^+ K^-)\times\mathcal{B}(X^+\to J/\psi\rho^+) < 4.2\times10^{-6}$.
Recently, Refs.\cite{wc0_1,wc0_2} predict that there must be isovector $D\bar{D}^*$ hadronic
molecules with $J^{PC}=1^{++}$ denotes as $W_{c1}^{0,\pm}$. They 
are isovector partner of $X(3872)$ in the hadronic molecular picture.
These states are virtual state poles at the $DD^*$ threshold, and appears as threshold cusps.  
The pole positions are 
$W^0_{c1}:~3881.7^{+1.0}_{-0.7}+i(1.2^{+0.8}_{-0.7})$ MeV and 
$W^{\pm}_{c1}:~3862.5^{+6.4}_{-10.3}-i(0.07)\pm0.00$ MeV.
The $W^0_{c1}$ pole is $10^{+1.0}_{-0.7}$ MeV above the $D^0\bar{D}^{*0}$
threshold and $1.8^{+1.0}_{-0.7}$ above the $D^+D^{*-}$ threshold.
The $W^-_{c1}$ pole is $13.3^{+10.3}_{-6.4}$ MeV below 
the $\bar{D}^0D^{*+}$ threshold.

The BESIII experiment has made seminal contributions to the study of the 
$\chi_{c1}(3872)$ in its decays, productions, and line shape, leveraging its unique capabilities in high-statistics data acquisition and low-background condition. This article gives a concise review 
and prospects of the radiative decays and line shape study of the $\chi_{c1}(3872)$ from the BESIII experiment.

\section{datasets at the BESIII experiment}
\begin{figure*}[!htbp]
\begin{center}
\includegraphics[width=0.6\textwidth]{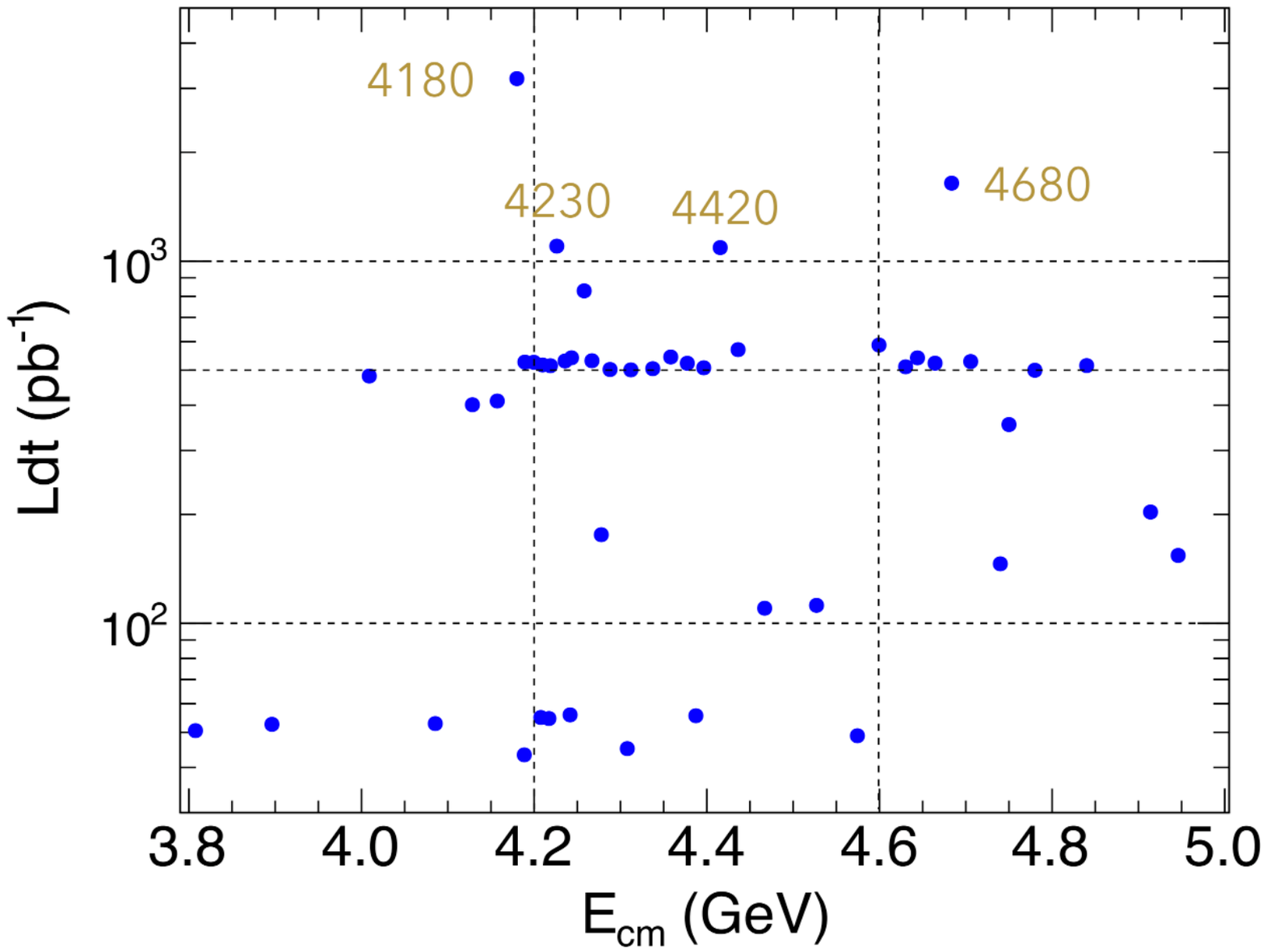}
\caption{
The luminosities of the datasets at center-of-mass energies above 3.8 GeV at BESIIII.
}
\label{fig_data}
\end{center}
\end{figure*}

BESIII at the BEPCII accelerator is a major upgrade of BESII at the BEPC for the studies of hadron physics and $\tau$-charm physics with the highest accuracy achieved until now.  In the past ten years, the BESIII experiment has accumulated a substantial 
dataset in the center-of-mass
energies above 3.8 GeV. Figure~\ref{fig_data} shows 
the luminosities at these energy points.
These data are used to investigate the physics on the charmonium-like states, charm mesons, and charmed baryons. BESIII has published a series of influential research outcomes including the discoveries of the $Z_c(3900)$ and $Z_{cs}(3885)$ using these data. 

\begin{figure*}[!htbp]
\includegraphics[width=0.46\textwidth]{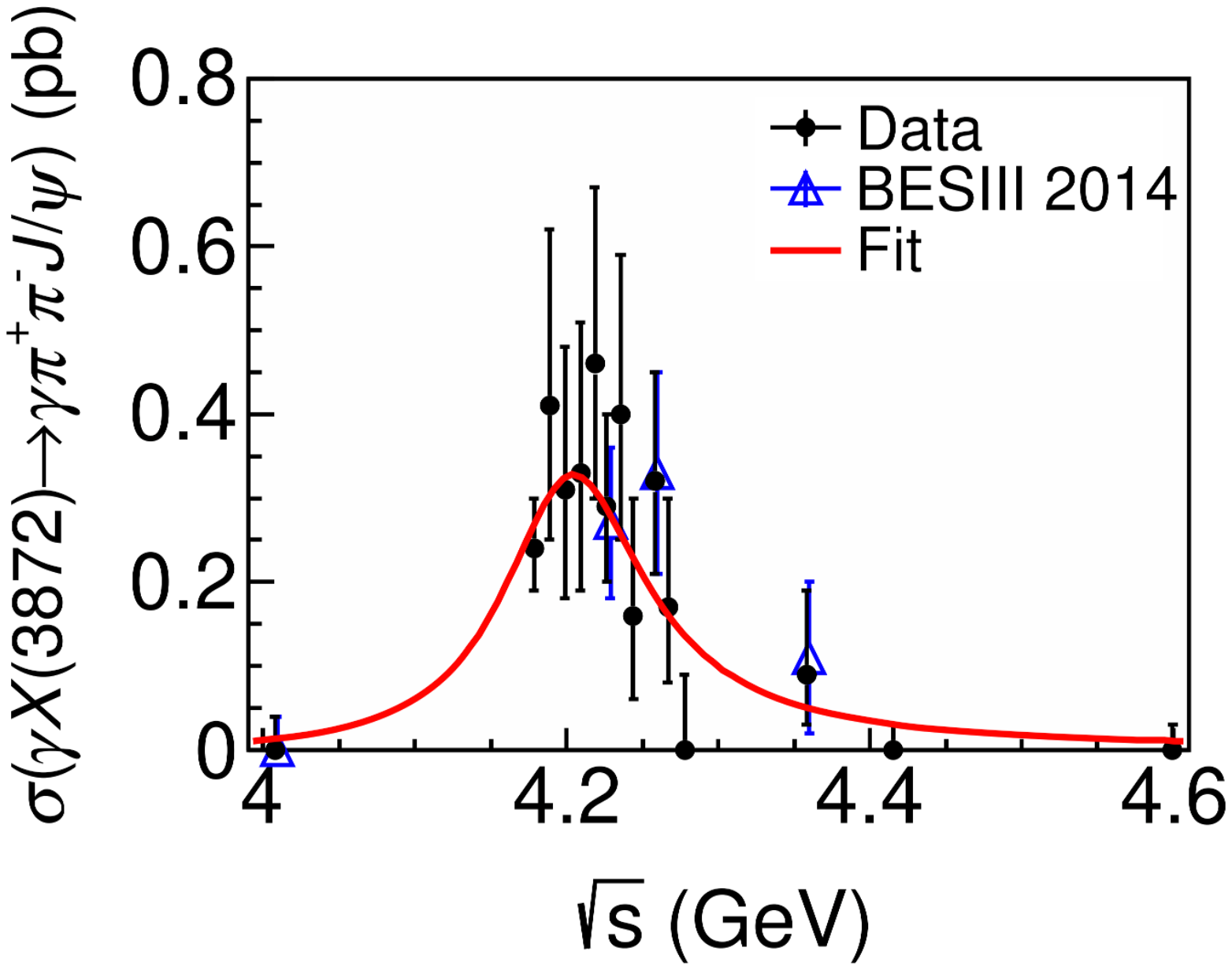}
\includegraphics[width=0.45\textwidth]{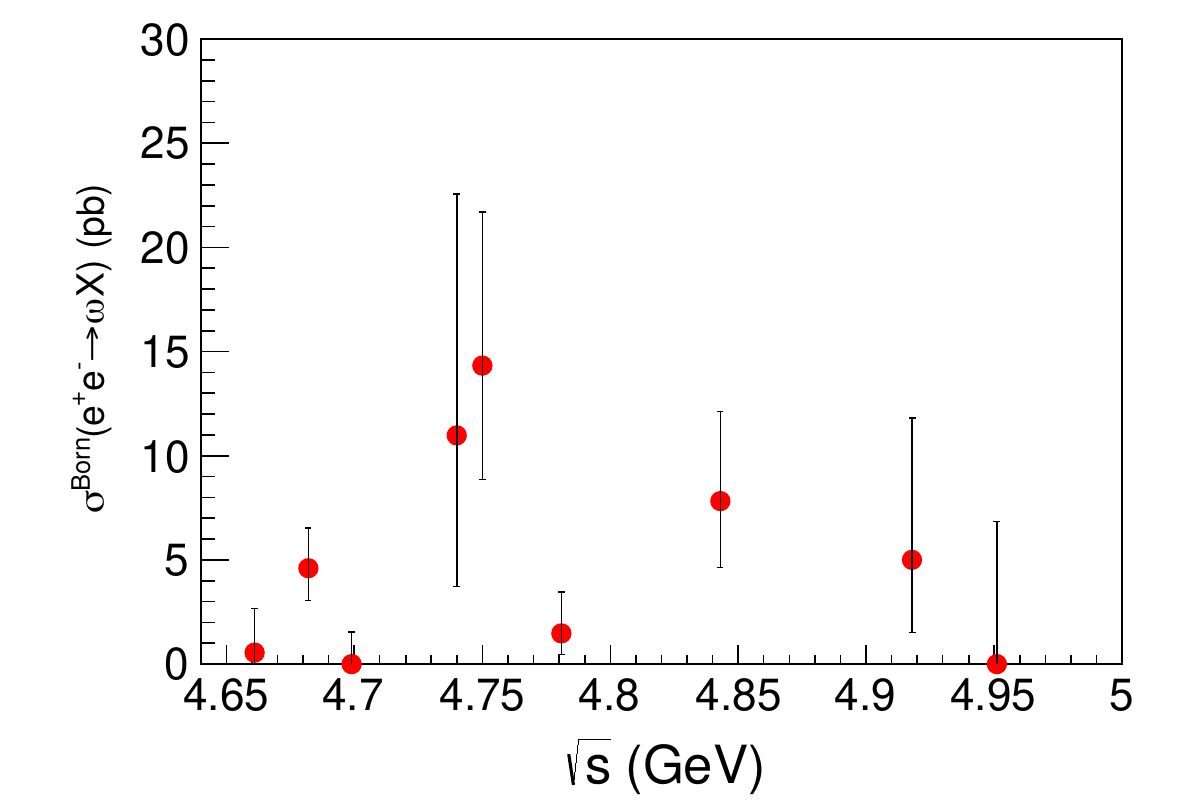}
\caption{
The $\sqrt{s}-$dependence cross sections of the processes $e^+e^-\to\gamma\chi_{c1}(3872)\to\gamma\pi^+\pi^-J/\psi$ (left)
and $e^+e^-\to\omega\chi_{c1}(3872)$ (right).
}
\label{fig_x}
\end{figure*}

BESIII made significant contributions to the investigation of the $\chi_{c1}(3872)$.
BESIII observed the $\chi_{c1}(3872)$ in the processes $e^+e^-\to\gamma\chi_{c1}(3872)$ at $\sqrt{s}$=4.18-4.3 GeV~\cite{Bs_g,Bs2} and $e^+e^-\to\omega\chi_{c1}(3872)$ above 4.66 GeV~\cite{Bs_omg}.
Figure~\ref{fig_x} shows the $\sqrt{s}-$dependence cross sections of the $e^+e^-\to\gamma\chi_{c1}(3872)\to\gamma\pi^+\pi^-J/\psi$ (left)
and $e^+e^-\to\omega\chi_{c1}(3872)$ (right).
Using these produced $\chi_{c1}(3872)$,  the decays and line shape of this particle are investigated at BESIII.

\section{$\chi_{c1}(3872)$ and $\psi(4230)$}

\begin{figure*}[!htbp]
\includegraphics[width=0.5\textwidth]{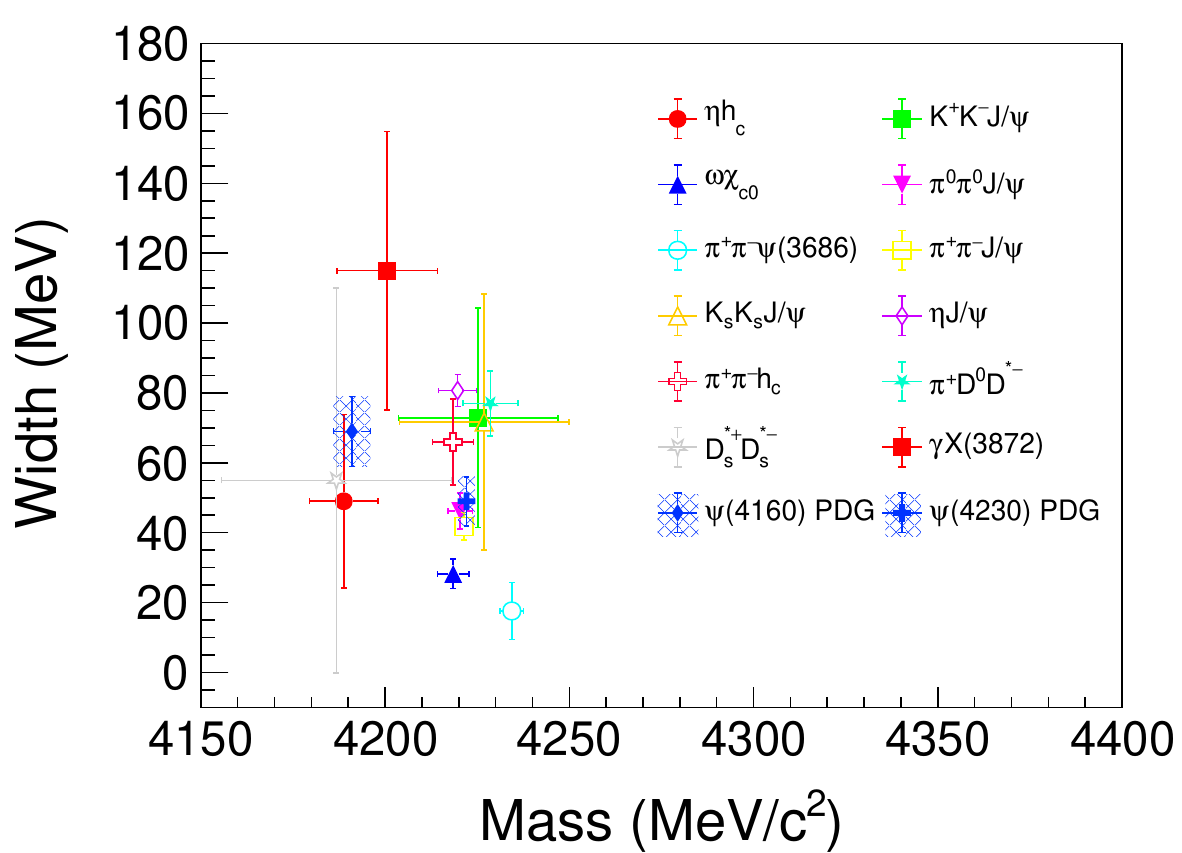}
\caption{
The $\psi(4230)$ mass and width measured in different processes.
The combined mass and width of the $\psi(4230)$
and $\psi(4160)$ provided by PDG are shown.
}
\label{fig_par}
\end{figure*}

As we mentioned before, one of the dominant way to produce the $\chi_{c1}(3872)$ at BESIII experiment is 
via the radiative process $e^+e^-\to\gamma\chi_{c1}(3872)$.
Benefit from the feature of the machine, BESIII 
could construct the relationship between the 
$\chi_{c1}(3872)$ and another famous vector charmonium-like state, $\psi(4230)$, which was 
observed firstly by the BaBar experiment.
Taking advantage of the scan data samples
collected by the detectors, 
BESIII measured the 
$\sqrt{s}-$dependence cross sections of
a series of $e^+e^-$ annihilation processes, and further reported the masses and widths 
of the structures observed in the line-shape of the cross section. Figure~\ref{fig_par} shows 
the measured values in the different processes.
The combined mass and width of the $\psi(4230)$
and $\psi(4160)$ provided in PDG~\cite{pdg} by globally 
analyzing the measurements from the different processes
are also shown in the Figure.
In addition, the $\sqrt{s}-$dependence cross sections of the process $e^+e^-\to\gamma\chi_{c1}(3872)$ via the decays $\chi_{c1}(3872)\to\pi^+\pi^-J/\psi$ and $\omega J/\psi$ are also measured by BESIII~\cite{Bs_g}, and 
the line-shape of which supports that the produced $\chi_{c1}(3872)$ is from the $\psi(4230)$ radiative
decay. By comparing the mass and width of  
the structure
measured via $e^+e^-\to\gamma\chi_{c1}(3872)$ to 
the global values of $\psi(4230)$ and $\psi(4160)$ as 
shown in Figure~\ref{fig_par}, we can see the discrepancies between these values. To explain 
this, Ref.~\cite{vec} suggests that the 
contributions to the $e^+e^-\to\gamma\chi_{c1}(3872)$
is not only from the $\psi(4230)$ but also the $\psi(4160)$. While a precise measurement of 
the line-shape of $e^+e^-\to\gamma\chi_{c1}(3872)$
cross section is essential to clarify 
the condition.

\section{The radiative decays of the $\chi_{c1}(3872)$}
\subsection{$\chi_{c1}(3872)\to\gamma J/\psi,\gamma\psi(2S)$}

The radiative decays of the $\chi_{c1}(3872)$ serve as crucial probes for elucidating the intrinsic properties of this particle. Independent measurements of these decays have been conducted by the BESIII, BaBar, Belle, and LHCb collaborations. 
Belle, BaBar, and LHCb all reported the measurements of $\chi_{c1}(3872)\to\pi^+\pi^-J/\psi, \gamma J/\psi, \gamma\psi(2S)$ channels in the $B^{\pm,0}\to\chi_{c1}(3872)K^{\pm,0}$ decays~\cite{Be2011a,Ba2008c,Be2011b,Ba2009b,lhcb2024,lhcb2020}. 
BESIII studied these channels via the $e^+e^-\to\gamma\chi_{c1}(3872)$ production process~\cite{flat_bes}.

\begin{figure*}[!htbp]
\includegraphics[width=0.4\textwidth]{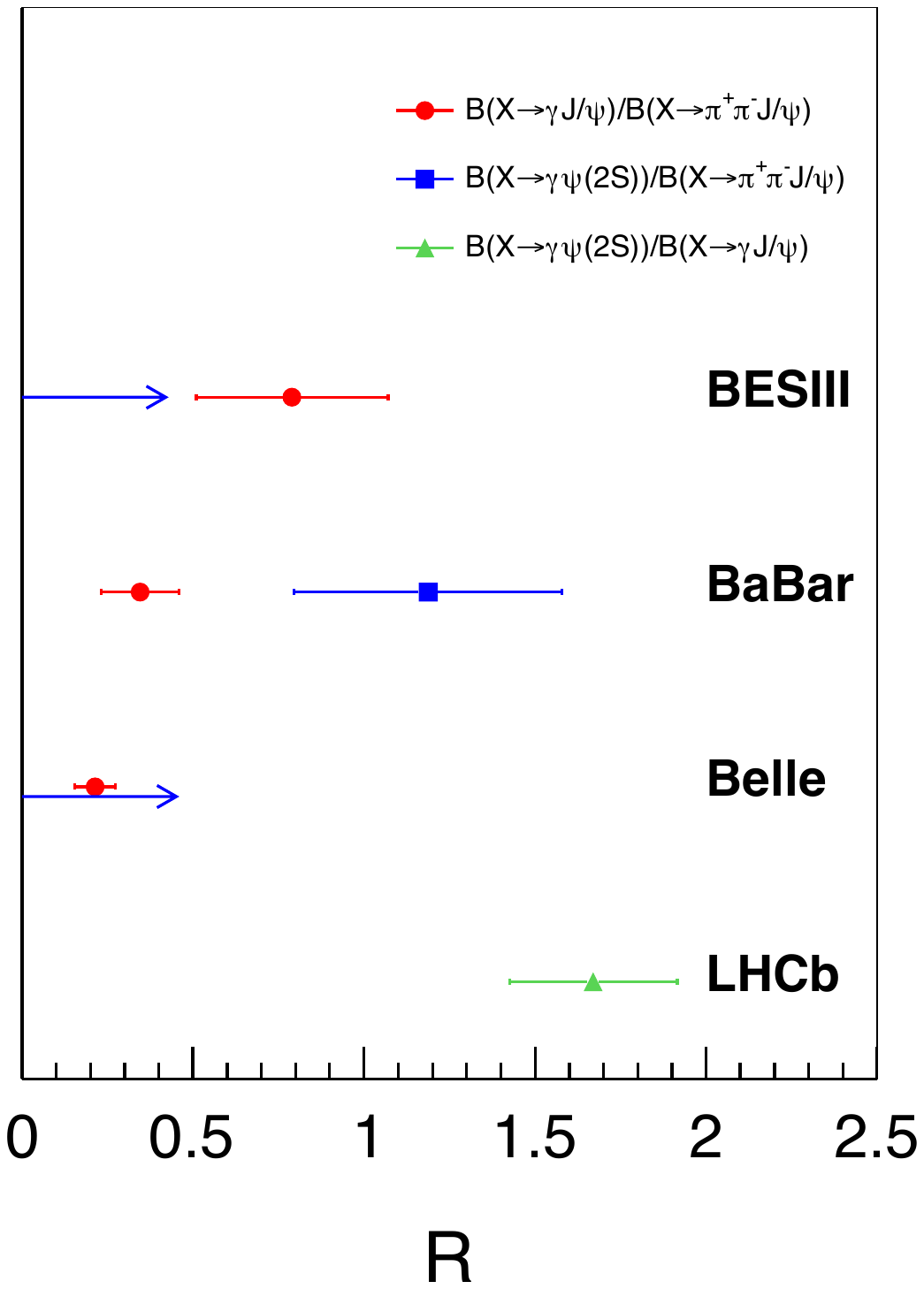}
\caption{
The branching fraction ratios measured by BESIII, BaBar, Belle, and LHCb.
}
\label{fig_ra}
\end{figure*}

The $\chi_{c1}(3872)\to\pi^+\pi^-J/\psi$
decay mode, characterized by its low-background conditions and small measurement uncertainties across experiments, is selected as the normalization channel to compute the branching fraction ratios:
$\mathcal{B}(\chi_{c1}(3872)\to\gamma J/\psi)/\mathcal{B}(\chi_{c1}(3872)\to\pi^+\pi^-J/\psi)$ and $\mathcal{B}(\chi_{c1}(3872)\to\gamma\psi(2S))/\mathcal{B}(\chi_{c1}(3872)\to\pi^+\pi^-J/\psi)$.
This approach enables to check consistencies of the measurements on the radiative decays among experiments.
Figure~\ref{fig_ra} shows the calculated branching fraction ratios from BESIII, BaBar, and Belle.
For the $\chi_{c1}(3872)\to\gamma\psi(2S)$ channel,
Belle and BESIII observed no significant signals; thus, their 90\% confidence level (CL) upper limits are denoted by arrows. It is worth pointing out that the denominator $\mathcal{B}(\chi_{c1}(3872)\to\pi^+\pi^-J/\psi)$
in each ratio corresponds to the measurement from the same experiment as the numerator, ensuring independence between results from different collaborations. Furthermore, LHCb directly measured the ratio 
$\mathcal{B}(\chi_{c1}(3872)\to\gamma\psi(2S))/\mathcal{B}(\chi_{c1}(3872)\to\gamma J/\psi)$, which is incorporated into Figure 1.

For the ratio $\mathcal{B}(\chi_{c1}(3872)\to\gamma J/\psi)/\mathcal{B}(\chi_{c1}(3872)\to\pi^+\pi^-J/\psi)$,
Belle provides the most precision measurement, and the ratios between Belle and BESIII are inconsitent with each other within one standard deviation. 
For the $\chi_{c1}(3872)\to\gamma J/\psi$ study at BESIII, the $J/\psi$ is reconstructed with the decays 
$J/\psi\to e^+e^-/\mu^+\mu^-$,
the dominant backgrounds are from the $e^+e^-\to(\gamma\gamma)e^+e^-$ and $(\gamma\gamma)\mu^+\mu^-$ processes. These backgrounds are very serious, and difficult to be suppressed in the study with an 
conventional method. Some techniques like machine learning may be helpful to improve the measurement.
For the ratio $\mathcal{B}(\chi_{c1}(3872)\to\gamma\psi(2S))/\mathcal{B}(\chi_{c1}(3872)\to\pi^+\pi^-J/\psi)$, both Belle and BESIII give the upper limits which contradict the measurement from BaBar as shown in
the Figure~\ref{fig_ra}. 
Although LHCb does not give the absolute branching fraction of 
$B\to\chi_{c1}(3872)K\to\gamma\psi(2S)K$, it does observe this decay, and report
the ratio $\mathcal{B}(\chi_{c1}(3872)\to\gamma\psi(2S))/\mathcal{B}(\chi_{c1}(3872)\to\gamma J/\psi)=1.67\pm0.21\pm0.12\pm0.04$.
At BESIII, the $\chi_{c1}(3872)\to\gamma\psi(2S)$ is studied
with the $\psi(2S)\to\pi^+\pi^-J/\psi, J/\psi\to e^+e^-/\mu^+\mu^-$ decay.
The background condition is clean in the analysis, but taking into account the branching fractions 
of $\psi(2S)\to\pi^+\pi^-J/\psi, J/\psi\to e^+e^-/\mu^+\mu^-$, the total efficiency is quite low.
The contradiction of the measurements on 
$\chi_{c1}(3872)\to\gamma\psi(2S)$ between these experiments exist for a long time. For BESIII, larger statistical data and refined reconstruction technique are both needed to improve the sensitivity on this channel.

\subsection{$\chi_{c1}(3872)\to\gamma\psi_2(3823)$}
\begin{figure*}[!htbp]
\includegraphics[width=0.5\textwidth]{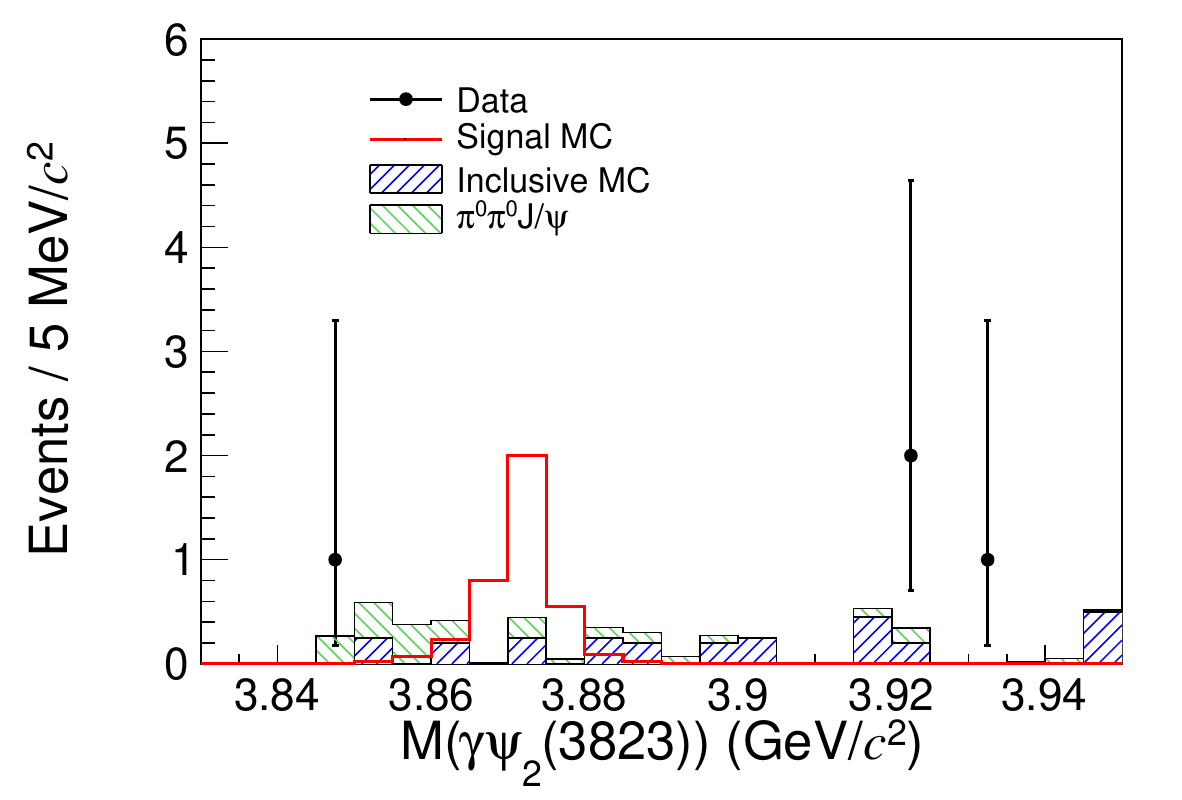}
\caption{
The distribution of the $\gamma\psi_2(3823)$ invariant mass at BESIII~\cite{cite_x3823}.
}
\label{x3823_ra}
\end{figure*}

Recently, BESIII reported the measurement of the 
radiative transition $\chi_{c1}(3872)\to\gamma\psi_2(3823)$ via the process $e^+e^-\to\gamma \chi_{c1}(3872)$ at center-of-mass energies $\sqrt{s}=4.178-4.278$ GeV~\cite{cite_x3823}.
Figure~\ref{x3823_ra} shows the obtained distribution of the $\gamma\psi_2(3823)$ invariant mass.
This is the first measurement on the $\chi_{c1}(3872)$ transition to a $D$-wave charmonium.
E1 transition is a natural decay between the $P$-wave an $D$-wave resonances and expected to be 
large decay width.
The radiative decay $\chi_{c1}(3872)\to\gamma \psi_2(3823)$ may happen via the E1 transition
if the $\chi_{c1}(3872)$ contains a component of the excited spin-triplet state $\chi_{c1}(2P)$, 
where the $\psi_2(3823)$ is considered as the $1^3D_2$ charmonium state. 
The branching fraction ratio of this decay relative to the 
$\chi_{c1}(3872)\to\pi^+\pi^-J/\psi$ decay, 
$\mathcal{R}_{\chi_{c1}(3872)}\equiv\mathcal{B}(\chi_{c1}(3872)\to\gamma \psi_2(3823), \psi_2(3823)\to\gamma\chi_{c1})/
\mathcal{B}(\chi_{c1}(3872)\to\pi^+\pi^- J/\psi)$, is investigated and the upper limit 
at the 90\% C.L. is set at 0.075 in this work.
Many theoretical models predict the partial widths 
of the radiative transitions between different conventional charmonium states including 
the $P$-wave an $D$-wave states.
The partial widths of $\chi_{c1}(2P)\to\gamma\psi(1^3D_2)$ and 
$\psi(1^3D_2)\to\gamma\chi_{c1}(1P)$
are calculated with the non-relativistic (NR) potential model
and the Godfrey-Isgur (GI) relativistic potential model~\cite{nr_gi}.
In addition, the lattice QCD (LQCD)~\cite{the_lqcd} calculated 
the partial width of $\psi(1^3D_2)\to\gamma\chi_{c1}(1P)$.
Incorporate with the total width of the $\psi(1^3D_2)$ estimated according to the BESIII measurements and some phenomenological results, the branching fraction ratio, 
$\mathcal{B}(\chi_{c1}(2P)\to\gamma\psi(1^3D_2),~\psi(1^3D_2)\to\gamma\chi_{c1}(1P))/\mathcal{B}(\chi_{c1}(3872)\to\pi^+\pi^- J/\psi)$, is estimated to be $0.46\pm0.19$, $0.21\pm0.09$, and $0.26\pm0.11$, based on 
NR, GI, and LQCD, respectively. The determined upper limit of 0.075 contradict these estimation, thus challenge the pure charmonium interpretation of $\chi_{c1}(3872)$.

BESIII implemented the $\chi_{c1}(3872)\to\gamma\psi_2(3823)$ study with a full reconstruction 
technique by reconstructing the $\psi_2(3823)$ with the cascade decay
$\psi_2(3823)\to\gamma\chi_{c1}$,
$\chi_{c1}\to\gamma J/\psi$,
$J/\psi\to \ell^+\ell^-$ ($\ell=e,\mu$), in which there are four radiative photons in the whole
decay chain.
In order to determine the order of the photons in the four class radiation, the mass constraints are applied to select the photons from the $\psi_2(3823)$ and $\chi_{c1}(1P)$ decays. This method is helpful to 
suppress the backgrounds and the combinatorial background in particular. The disadvantage of the approach, however, is to constrain the involved intermediate particles to be the masses of $\psi_2(3823)$ and $\chi_{c1}(1P)$, which limited the physical aims of the project. In principle, the project could be implemented with a more ambitious algorithm to search for all possible intermediate states, e.g. $\chi_{c0,1,2}(2P)$ and $\psi(1D)$, in the transitions $e^+e^-\to\gamma\chi_{c0,1,2}(2P)$, $\chi_{c0,1,2}(2P)\to\gamma\psi(1D)$, $\psi(1D)\to\gamma\chi_{c0,1,2}(1P)$, $\chi_{c0,1,2}(1P)\to\gamma J/\psi$.
Of course, a detailed study is needed to deal with the ranking problem of the four radiative photons.
Anyway, there is still space at BESIII to improve the sensitivity and provide more abundant physical results
for this project.

\section{The $\chi_{c1}(3872)$ line shape}
The extremely approximation of its mass to the $D^{*0}\bar{D}$ mass threshold makes the $\chi_{c1}(3872)$ line shape
reveals distinctly asymmetric behavior. It is promising to break through the puzzle of the $\chi_{c1}(3872)$ nature in
the investigation of its line shape. A Flatté-inpired parametrization is widely applied to study its line shape.  
In 2024, BESIII reported the coupled-channel analysis of the line shape on 
the channels $\chi_{c1}(3872)\to D^0\bar{D}^0\pi^0$ and $\pi^+\pi^-J/\psi$~\cite{flat_bes}. 
The parameters of the Flatté-inspired parametrization, e.g. the coupling constant, 
mass, and the decay width, are determined. The authors estimated the scattering 
length the effective range, and the results reveal the similar 
compositeness to that of deuteron, but the uncertainties are quite large and definitive conclusions need more data.
In the $\chi_{c1}(3872)\to D^0\bar{D}^0\pi^0$ spectrum, BESIII yields about 25 signals using the double tag charm mesons technique. The reconstruction algorithm could be tuned to improve the efficiency. Of course, a larger statistics data set is always essential to achieve precision measurement.

\section{Summary}
As a dedicated electron-positron collider experiment operating at the $\tau$-charm energy region, the BESIII  possesses unique advantages in studying the \(\chi_{c1}(3872)\). Its optimized detector geometry and high luminosity at center-of-mass energies near 3.8–5.0 GeV enable threshold-scanning precision for charmonium-like states. With the data set accumulated in the past decade, BESIII provided the measurements of the 
$\chi_{c1}(3872)\to\gamma J/\psi, \gamma\psi(2S), \gamma\psi_2(33823)$, $\bar{D}^{*0}D^0, \omega J/\psi, \pi^0\chi_{c1}(1P)$ decays, and 
discovered $\psi(4230)\to\gamma\chi_{c1}(3872)$ and $e^+e^-\to\omega\chi_{c1}(3872)$.
The coupled-channel analysis on the 
$\chi_{c1}(3872)\to\pi^+\pi^-J/\psi$ and $\bar{D}^{*0}D^0$ decays are implemented.
Building upon the inherent advantages of the BESIII experiment, the opportunities exist to further refine its event reconstruction capabilities, which could substantially enhance precision in exotic hadron studies. 
An imminent opportunity arises from the upcoming upgrade of the BEPCII collider, which will deliver a threefold enhancement in instantaneous luminosity within the XYZ energy region. This luminosity gain is projected to enable the collection of data samples with statistical significance far exceeding current levels across these critical energy points. Such a leap in data volume will provide an essential foundation for high-precision investigations of the \(\chi_{c1}(3872)\) at BESIII.

\end{document}